# The problem of reconstruction for static spherically-symmetric $4d$ metrics in scalar-Einstein-Gauss-Bonnet model

K. K. Ernazarov[1] and V. D. Ivashchuk[1,2]

[1] *Institute of Gravitation and Cosmology,*
*Peoples' Friendship University of Russia (RUDN University),*
*6 Miklukho-Maklaya Street, Moscow, 117198, Russian Federation,*

[2] *Center for Gravitation and Fundamental Metrology,*
*Scientific Research Center of Applied Metrology Rostest,*
*46 Ozyornaya Street, Moscow, 119361, Russian Federation.*

**Abstract**

We consider the $4d$ gravitational model with scalar field $\varphi$, Einstein and Gauss-Bonnet terms. The action of the model contains potential term $U(\varphi)$, Gauss-Bonnet coupling function $f(\varphi)$ and parameter $\varepsilon = \pm 1$, where $\varepsilon = 1$ corresponds to ordinary scalar field and $\varepsilon = -1$ - to phantom one. Inspired by recent works of Nojiri and Nashed, we explore a reconstruction procedure for a generic static spherically symmetric metric written in Buchdal parametrisation: $ds^2 = (A(u))^{-1} du^2 - A(u) dt^2 + C(u) d\Omega^2$, with given $A(u) > 0$ and $C(u) > 0$. The procedure gives the relations for $U(\varphi(u))$, $f(\varphi(u))$ and $d\varphi/du$, which lead to exact solutions to equations of motion with a given metric. A key role in the approach play the solutions to a second order linear differential equation for the function $f(\varphi(u))$. The formalism is illustrated by two examples when: a) Schwarzschild metric and b) Ellis wormhole metric, are chosen as a starting point. For the first case a) the black hole solution with "trapped ghost" is found which describes an ordinary scalar field outside the photon sphere and phantom scalar field inside the photon sphere ("trapped ghost"). For the second case b) the sEGB-extension of the Ellis wormhole solution is found when the coupling function reads: $f(\varphi) = c_1 + c_0(\tan(\varphi) + \frac{1}{3}(\tan(\varphi))^3)$, where $c_1$ and $c_0$ are constants.

# 1 Introduction

For decades, theoretical physicists have been working to unify gravity with quantum mechanics. String theory, once hailed as a promising candidate for such unification, "predicted" the existence of hidden extra dimensions and a variety of new fields, including the scalar dilaton. In the low energy limit, string theory also predicts certain extensions of General Relativity (GR). A possible extension involves incorporating the Gauss-Bonnet (GB) term [1, 2, 3, 4], coupled to a certain function of a scalar field (dilaton), giving rise to rather rich, but complex, landscape of scalar-Einstein-Gauss-Bonnet (sEGB) gravity.

It is worth noting that the pure GB term yields a topological invariant in four dimensions, while it becomes dynamically relevant in higher dimensions. The emergence of sEGB gravity challenges the standard black hole paradigm established by GR. A variety of non-trivial couplings between the scalar field and the GB term induces certain deviations from the Schwarzschild solution, ushering in a new "trend" of "hairy" black holes, which are characterized by scalar "hairs". The so-called scalarization, studied extensively by Kanti et al. [5, 6] and others, has profound implications for black hole properties, influencing their computability, stability, thermodynamics, and interaction with surrounding matter, see [7, 8] and references therein. Some authors, e.g. J. Kunz et al. and others, have extensively investigated static and rotating black hole solutions for certain models, revealing their unique characteristics [8, 9]. These black holes have a new parameter - scalar charge, which affects their gravitational and thermodynamic properties [10].

A significant advance in understanding black hole solutions within the scalar-Einstein-Gauss-Bonnet (sEGB) model was achieved by Bronnikov and Elizalde [11]. Their work demonstrated that the Gauss-Bonnet term circumvents established "no-go" theorems, which strictly limit the existence of certain black hole configurations in general relativity with a minimally coupled scalar field. This finding opens up new possibilities for the theoretical exploration of black hole properties within the sEGB framework.

The theoretical underpinnings of sEGB gravity are compelling, but observational evidence is crucial to validate its predictions. Fortunately, sEGB black holes possess distinct observational signatures detectable through various astrophysical probes. Gravitational waves are one such probe; merging black holes in sEGB gravity should emit gravitational waves exhibiting characteristic deviations from General Relativity (GR) predictions. Recent detection and analysis of these gravitational waves by detectors, such as LIGO and Virgo, provide a powerful means of testing sEGB gravity and constraining its model parameters [12].

The shadows [13] and quasinormal modes [14] of sEGB black holes offer promising avenues for investigation. A black hole's shadow, which is its dark silhouette against a bright background, is shaped by its geometry and surrounding spacetime. Cunha et al. [7] demonstrated that sEGB black holes possess distinctive, non-circular shadow morphologies, unlike the circular shadows predicted by General Relativity (GR). Furthermore, the quasinormal modes, which are characteristic frequencies emitted during perturbations, are sensitive to the presence of the scalar field and Gauss-Bonnet (GB) term [16]. These observational features provide unique ways to distinguish sEGB black holes from their GR counterparts.

Inspired by recent articles by Nojiri and Nashed [17, 18] on static spherically-symmetric solutions within the sEGB model and its fluid extension, this paper investigates the model governed by coupling function $f(\varphi)$ and potential function $U(\varphi)$, where $\varphi$ is a scalar field, either ordinary or phantom one.

## 2 The scalar-Einstein-Gauss-Bonnet model

Here we deal with the so-called scalar-Einstein-Gauss-Bonnet model which is governed by the action

$$S = \int d^4z \, |g|^{\frac{1}{2}} \left( \frac{R(g)}{2\kappa^2} - \frac{1}{2}\varepsilon g^{MN}\partial_M\varphi\partial_N\varphi - U(\varphi) + f(\varphi)\mathcal{G} \right), \tag{2.1}$$

where $\kappa^2 = 8\pi\frac{G_N}{c^4}$ ($G_N$ is Newton's gravitational constant, $c$ is speed of light), $\varphi$ is scalar field, $g_{MN}dz^Mdz^N$ is $4d$ metric of signature $(-,+,+,+)$, $R[g]$ is scalar curvature, $\mathcal{G}$ is Gauss-Bonnet term, $U(\varphi)$ is potential, $f(\varphi)$ is coupling function and $\varepsilon = \pm 1$.

For ordinary scalar field we have $\varepsilon = 1$, while for phantom one we should put $\varepsilon = -1$.

We study the spherically-symmetric solutions with the metric

$$ds^2 = g_{MN}(z)dz^M dz^N = e^{2\gamma(u)}du^2 - e^{2\alpha(u)}dt^2 + e^{2\beta(u)}d\Omega^2 \tag{2.2}$$

defined on the manifold

$$M = \mathbb{R} \times \mathbb{R}_* \times S^2. \tag{2.3}$$

Here $\mathbb{R}_* = (2\mu, +\infty)$ and $S^2$ is 2-dimensional sphere with the metric $d\Omega^2 = d\theta^2 + sin^2\theta d\varphi^2$, where $0 < \theta < \pi$ and $0 < \varphi < 2\pi$.

By substitution the metric (2.2) into the action we obtain

$$S = 4\pi \int du \left( L + \frac{dF_*}{du} \right), \tag{2.4}$$

where the Lagrangian $L$ reads

$$L = \frac{1}{\kappa^2} \left( e^{\alpha-\gamma+2\beta} \dot{\beta} \left( \dot{\beta} + 2\dot{\alpha} \right) + e^{\alpha+\gamma} \right)$$
$$-\frac{1}{2} e^{\alpha-\gamma+2\beta} \varepsilon \dot{\varphi}^2 - e^{\alpha+\gamma+2\beta} U(\varphi) - 8\dot{\alpha}\dot{\varphi} \frac{df}{d\varphi} \left( \dot{\beta}^2 e^{\alpha+2\beta-3\gamma} - e^{\alpha-\gamma} \right), \tag{2.5}$$

and the total derivative term is irrelevant for our consideration. Here and in what follows we denote $\dot{x} = \frac{dx}{du}$. The relation (2.4) is derived in Appendix, where explicit relation for the $F_*$ term is given.

The equations of motion for the action (2.1) with the metric (2.2) involved are equivalent to Lagrange equation corresponding to the Lagrangian (2.5), which have the following form[1]

$$\frac{\partial L}{\partial \gamma} = \frac{1}{\kappa^2} \left( -e^{\alpha-\gamma+2\beta} \dot{\beta} \left( \dot{\beta} + 2\dot{\alpha} \right) + e^{\alpha+\gamma} \right) + \frac{1}{2} e^{\alpha-\gamma+2\beta} \varepsilon \dot{\varphi}^2 - e^{\alpha+\gamma+2\beta} U(\varphi)$$
$$-8\dot{\alpha}\dot{\varphi} \frac{df}{d\varphi} \left( -3\dot{\beta}^2 e^{\alpha+2\beta-3\gamma} + e^{\alpha-\gamma} \right) = 0, \tag{2.6}$$

$$\frac{d}{du}\left(\frac{\partial L}{\partial \dot{\alpha}}\right) - \frac{\partial L}{\partial \alpha} = \frac{d}{du} \left( \frac{2}{\kappa^2} e^{\alpha-\gamma+2\beta} \dot{\beta} - 8\dot{\varphi} \frac{df}{d\varphi} \left( \dot{\beta}^2 e^{\alpha+2\beta-3\gamma} - e^{\alpha-\gamma} \right) \right)$$
$$- \left( \frac{1}{\kappa^2} \left( e^{\alpha-\gamma+2\beta} \dot{\beta} \left( \dot{\beta} + 2\dot{\alpha} \right) + e^{\alpha+\gamma} \right) - \frac{1}{2} e^{\alpha-\gamma+2\beta} \varepsilon \dot{\varphi}^2 - e^{\alpha+\gamma+2\beta} U(\varphi) \right. \tag{2.7}$$
$$\left. -8\dot{\alpha}\dot{\varphi} \frac{df}{d\varphi} \left( \dot{\beta}^2 e^{\alpha+2\beta-3\gamma} - e^{\alpha-\gamma} \right) \right) = 0,$$

---
[1] Here by relations (2.6) and (2.7) we eliminate the typos in relations (2.5) and (2.6) from Ref. [19].

$$\frac{d}{du}\left(\frac{\partial L}{\partial \dot{\beta}}\right) - \frac{\partial L}{\partial \beta} = \frac{d}{du}\left(\frac{1}{\kappa^2}e^{\alpha-\gamma+2\beta}\left(2\dot{\beta}+2\dot{\alpha}\right) - 8\dot{\alpha}\dot{\varphi}\frac{df}{d\varphi}2\dot{\beta}e^{\alpha+2\beta-3\gamma}\right)$$

$$-\left(\frac{2}{\kappa^2}e^{\alpha-\gamma+2\beta}\dot{\beta}\left(\dot{\beta}+2\dot{\alpha}\right) - e^{\alpha-\gamma+2\beta}\varepsilon\dot{\varphi}^2 - 2e^{\alpha+\gamma+2\beta}U\left(\varphi\right)\right. \tag{2.8}$$

$$\left. -16\dot{\alpha}\dot{\varphi}\frac{df}{d\varphi}\dot{\beta}^2 e^{\alpha+2\beta-3\gamma}\right) = 0,$$

and

$$\frac{d}{du}\left(\frac{\partial L}{\partial \dot{\varphi}}\right) - \frac{\partial L}{\partial \varphi} = \frac{d}{du}\left(-e^{\alpha-\gamma+2\beta}\varepsilon\dot{\varphi} - 8\dot{\alpha}\frac{df}{d\varphi}\left(\dot{\beta}^2 e^{\alpha+2\beta-3\gamma} - e^{\alpha-\gamma}\right)\right)$$

$$-\left(-e^{\alpha+\gamma+2\beta}\frac{dU}{d\varphi} - 8\dot{\alpha}\dot{\varphi}\frac{d^2f}{d\varphi^2}\left(\dot{\beta}^2 e^{\alpha+2\beta-3\gamma} - e^{\alpha-\gamma}\right)\right) = 0. \tag{2.9}$$

# 3 Reconstruction procedure

Here we use (without loss of generality) the Buchdal radial gauge obeying $\alpha = -\gamma$. For the metric (2.2) we obtain

$$ds^2 = \left(A\left(u\right)\right)^{-1}du^2 - A\left(u\right)dt^2 + C\left(u\right)d\Omega^2, \tag{3.1}$$

where

$$e^{2\gamma(u)} = \left(A\left(u\right)\right)^{-1}, \quad e^{2\alpha(u)} = A\left(u\right) > 0, \quad e^{2\beta(u)} = C\left(u\right) > 0. \tag{3.2}$$

In what follows we use the identities

$$\dot{\alpha} = \frac{\dot{A}}{2A}, \qquad \dot{\beta} = \frac{\dot{C}}{2C}. \tag{3.3}$$

We put (without loss of generality) $\kappa^2 = 1$. We also denote

$$f\left(\varphi\left(u\right)\right) = f, \qquad U\left(\varphi\left(u\right)\right) = U \tag{3.4}$$

and hence

$$\frac{d}{du}f = \frac{df}{d\varphi}\frac{d\varphi}{du} \iff \dot{f} = \frac{df}{d\varphi}\dot{\varphi}, \tag{3.5}$$

$$\frac{d}{du}U = \frac{dU}{d\varphi}\frac{d\varphi}{du} \iff \dot{U} = \frac{dU}{d\varphi}\dot{\varphi}. \tag{3.6}$$

Multiplying (2.6) by $(-2)$ and using relations (3.2), (3.3), (3.5) we get

$$\dot{A}\left[8\dot{f}\left(1-3KA\right)+\dot{C}\right] + 2KA - 2 - CA\varepsilon\dot{\varphi}^2 + 2CU = 0, \tag{3.7}$$

where here and in what follows we use the notation

$$K \equiv \left(\frac{\dot{C}}{2C}\right)^2 C. \tag{3.8}$$

We note, that for $C(u) = u^2$ we have $K = 1$ and the equation (3.7) is coinciding with the equation (10) from Ref. [17].

Multiplying (2.7) by 2 and using relations (3.2), (3.3), (3.5), (3.8) we get

$$16\ddot{f}A(1 - KA) + 8\dot{f}\left(\dot{A} - 3KA\dot{A} - 2\dot{K}A^2\right) + \dot{A}\dot{C}$$
$$+ 2A\left(\ddot{C} - K\right) + CA\varepsilon\dot{\varphi}^2 - 2 + 2CU = 0. \tag{3.9}$$

In special case $C(u) = u^2$ the equation (3.9) is coinciding with the equation (9) from Ref. [17].

Analogously, using (3.2), (3.3), (3.5), (3.8) we rewrite equation (2.8) as follows

$$\left(C - 4\dot{f}A\dot{C}\right)\ddot{A} - 4\ddot{f}\dot{A}A\dot{C} - 4\dot{f}\left(\left(\dot{A}\right)^2\dot{C} + \dot{A}A\ddot{C} - 2\dot{A}AK\right) +$$
$$+ \dot{A}\dot{C} + C\left(A\varepsilon\dot{\varphi}^2 + 2U\right) + A\ddot{C} - 2AK = 0. \tag{3.10}$$

Notice, that for $C(u) = u^2$ the equation (3.9) is coinciding with the equation (11) from Ref. [17].

Now, multiplying equation (2.9) by $(-\dot{\varphi})$ we obtain

$$4\dot{f}(AK - 1)\ddot{A} + \varepsilon\ddot{\varphi}\dot{\varphi}AC + 4\dot{f}\dot{A}\left(\dot{A}K + A\dot{K}\right)$$
$$+ \left(\dot{A}C + A\dot{C}\right)\varepsilon\dot{\varphi}^2 - C\dot{U} = 0. \tag{3.11}$$

For $C(u) = u^2$ the equation (3.11) is coinciding with the equation (12) from Ref. [17].

Here we put the following restriction

$$\dot{\varphi} \neq 0 \quad \text{for} \quad u \in (u_-, u_+), \tag{3.12}$$

where interval $(u_-, u_+)$ is belonging to $\mathbb{R}_* = (2\mu, +\infty)$. Then the relations (3.11) and (2.9) are equivalent in this interval.

By adding equations (3.9) and (3.7) and dividing the result by $4C$ we get the relation for the function $U = U(\varphi(u))$

$$U = \frac{1}{C}\left(1 - 4A(1 - KA)\ddot{f} - \dot{A}\left[4\dot{f}(1 - 3KA) + \frac{1}{2}\dot{C}\right] + 4\dot{f}KA^2 - \frac{1}{2}A\ddot{C}\right). \tag{3.13}$$

For special choice $C(u) = u^2$ and $\varepsilon = 1$ the equation (3.13) is coinciding with the equation (3.12) from Ref. [19].

The relation (3.13) may be written as

$$CU = E_U\ddot{f} + F_U\dot{f} + G_U, \tag{3.14}$$

where
$$E_U = -4A(1 - KA), \tag{3.15}$$
$$F_U = -4\dot{A}(1 - 3KA) + 4\dot{K}A^2, \tag{3.16}$$
$$G_U = 1 - \frac{1}{2}\dot{A}\dot{C} - \frac{1}{2}A\ddot{C}. \tag{3.17}$$

Subtracting (3.7) from (3.9) and dividing the result by $2A$, we obtain the relation for $\dot{\varphi}$
$$C\varepsilon\dot{\varphi}^2 = 8\ddot{f}(KA - 1) + 8\dot{f}\dot{K}A + 2K - \ddot{C} \equiv \Phi_\varepsilon. \tag{3.18}$$

In the special case $C(u) = u^2$ and $\varepsilon = 1$ agreement with the relation (14) from Ref. [17].

Due to $C(u) > 0$ and (3.12) we get
$$\varepsilon\Phi_\varepsilon > 0 \tag{3.19}$$

for all $u \in (u_-, u_+)$.

In the simplest case $C(u) = u^2$ and $\varepsilon = 1$ explored in Ref. [17] we get $K = 1$, $\ddot{C} = 2$ and hence the restriction (3.19) reads
$$8\ddot{f}(A - 1) = \Phi_1/u^2 > 0. \tag{3.20}$$

Subtracting (3.9) from (3.10), we get the master equation for the function $f = f(\varphi(u))$
$$E\ddot{f} + F\dot{f} + G = 0, \tag{3.21}$$

where
$$E = 4A\left(4KA - \dot{A}\dot{C} - 4\right), \tag{3.22}$$
$$F = -4\ddot{A}A\dot{C} - 4\left(\dot{A}\right)^2\dot{C} - 4\dot{A}A\ddot{C} + 8\left(4KA\dot{A} - \dot{A} + 2\dot{K}A^2\right), \tag{3.23}$$
$$G = C\ddot{A} - A\ddot{C} + 2. \tag{3.24}$$

For $\varepsilon = 1$ and $C(u) = u^2$ the master equation (3.21) is coinciding with equation (15) from Ref. [17].

**Validation of equations of motion.** The set of obtained equations: (3.13), (3.18), (3.21) is equivalent to set of first three equations of motion: (2.6), (2.7), (2.8). Indeed, due to construction equations (2.6), (2.7), (2.8) are equivalent to equations (3.7), (3.9), (3.10), respectively, which may be written as $X_1 = 0$, $X_2 = 0$, $X_3 = 0$. Then, due to construction equations (3.13), (3.18), (3.21) may be written as $Y_1 = X_1 + X_2 = 0$, $Y_2 = X_1 - X_2 = 0$, $Y_3 = X_3 - X_2 = 0$. It is obvious that the latter set of equations: $Y_1 = 0$, $Y_2 = 0$, $Y_3 = 0$ is equivalent to the former one: $X_1 = 0$, $X_2 = 0$, $X_3 = 0$. Hence the set of equations (3.13), (3.18), (3.21) is equivalent to the set of equations (3.7), (3.9), (3.10), which is equivalent to the set of equations (2.6), (2.7), (2.8). As to the equation of motion corresponding to scalar field (2.9) it follows just from the obtained equations: (3.13), (3.18), (3.21), when condition (3.12) is obeyed. Indeed, due to $\dot{\varphi} \neq 0$ the equation (2.9) is equivalent to the equation (3.11). The equation (3.11) may be verified by differentiation of equations (3.13) and (3.18) with respect to $u$, substituting obtained relations for $\varepsilon\ddot{\varphi}\dot{\varphi}$ and $\dot{U}$, and also for $\varepsilon\dot{\varphi}^2$ into the left hand side of (3.11) and using master equation (3.21).

# 4 Solutions to master eqution

Here we consider the solutions to master equation (3.21).

## 4.1 The case $E \neq 0$

First, we put
$$E(u) \neq 0 \quad \text{for} \quad u \in (u_-, u_+), \tag{4.1}$$

where $(u_-, u_+)$ is interval from (3.12). Denoting $y = \dot{f}$ we rewrite equation (3.21) as

$$\dot{y} + a(u) y + b(u) = 0, \tag{4.2}$$

where

$$a(u) = \frac{F(u)}{E(u)}, \qquad b(u) = \frac{G(u)}{E(u)}. \tag{4.3}$$

The solution to differential equation (4.2) can be readily obtained by using standard methods. This solution reads

$$\dot{f} = y = C_0 y_0(u) - y_0(u) \int_{u_0}^{u} dw\, b(w) (y_0(w))^{-1}, \tag{4.4}$$

where $u \in (u_-, u_+)$, $C_0$ is constant and

$$y_0(u) = \exp\left(-\int_{u_0}^{u} dv\, a(v)\right) \tag{4.5}$$

is the solution to homogeheous linear differential equation: $\dot{y}_0 + a(u) y_0 = 0$.

Integrating (4.4) we obtain

$$f = C_1 + C_0 \int_{u_0}^{u} dv\, y_0(v) - \int_{u_0}^{u} dv\, y_0(v) \int_{u_0}^{v} dw\, b(w) (y_0(w))^{-1}, \tag{4.6}$$

where $C_1$ is a constant. We note that relation (3.19) impose restrictions only on $C_0$ and $u_0$ since the function $\Phi_\varepsilon$ depends on $\dot{f}$ and $\ddot{f}$. Here $C_1$ is an arbitrary constant.

## 4.2 The case $E = 0$

Now we put
$$E(u) = 0 \quad \text{for all} \quad u \in (u_-, u_+). \tag{4.7}$$

In this case the master equation (3.21) reads

$$F\dot{f} + G = 0. \tag{4.8}$$

In the subcase

$$F(u) \neq 0 \quad \text{for all} \quad u \in (u_-, u_+) \tag{4.9}$$

the solution to master equation (4.8) has the following form

$$f = C_1 - \int_{u_0}^{u} dv d(v), \qquad d(v) = \frac{G(v)}{F(v)}, \tag{4.10}$$

where $C_1$ is a constant.

For the subcase, when

$$F(u) = E(u) = 0 \quad \text{for all} \quad u \in (u_-, u_+), \tag{4.11}$$

the master equation has a solution only if

$$G(u) = 0 \quad \text{for all} \quad u \in (u_-, u_+). \tag{4.12}$$

In this case function $f$ is arbitrary.

It may be verified that for the functions $E(u), F(u), G(u)$ defined in (3.22), (3.23), (3.24), respectively, the relations (4.11) imply (4.12) when $A(u) \neq 0$. This means that the only "dangerous" case (of the absence of solutions to master equation) when $F(u) = E(u) = 0$ for all $u \in (u_-, u_+)$ and $G(u) \neq 0$ for some $u \in (u_-, u_+)$ does not take place.

Indeed, by plugging the definitions (3.22) and (3.23) for the functions $E(u)$ and $F(u)$ into equations (4.11) we obtain (e.g. by Maple) two sets of functions $A(u), C(u)$:

$$(i) \ C(u) = (d_1 u + d_0)^2, \qquad A(u) = \frac{1}{d_1^2}, \quad d_1 \neq 0, \tag{4.13}$$

$$(ii) \quad C(u) \text{ is arbitrary}, \qquad A(u) = 0. \tag{4.14}$$

The second set $(ii)$ is excluded by the restriction $A(u) \neq 0$. For the first case $(i)$ we get $G(u) = 0$ from definition (3.24). It should be noted that the set $(i)$ after substitution into the metric (3.1) leads us (under an appropriate reparametrization of $t$ and $u$ variables) just to Minkowski metric.

## 5 Examples

Here we consider two examples of reconstruction procedure to test the reconstruction scheme under consideration.

### 5.1 Schwarzschild metric

Let us start with the simplest case of the Schwarzschild metric with

$$A(u) = 1 - \frac{2\mu}{u}, \quad C(u) = u^2, \tag{5.1}$$

where $\mu > 0$ and $u > 2\mu$. In this case for the master equation (3.21) we get the following relations for the functions $E(u)$, $F(u)$ and $G(u)$ defined in (3.22), (3.23) and (3.24), respectively:

$$E = -\frac{48\mu}{u^2}(u - 2\mu), \tag{5.2}$$

$$F = \frac{64\mu}{u^3}(u - 3\mu), \tag{5.3}$$

$$G = 0. \tag{5.4}$$

Solving the master equation $E\ddot{f} + F\dot{f} + G = 0$ we obtain for coupling function

$$f = f(\varphi(u)) = c_1 + c_0 \frac{3}{7}(u - 2\mu)^{\frac{1}{3}}\left(u^2 + 3\mu u + 18\mu^2\right) \tag{5.5}$$

and

$$\dot{f} = c_0 u^2 (u - 2\mu)^{-\frac{2}{3}}, \quad \ddot{f} = c_0 \frac{4u(u - 3\mu)}{3(u - 2\mu)^{\frac{5}{3}}}, \tag{5.6}$$

where $c_0$ and $c_1$ are constants and $u > 2\mu$. (Here the integration constants in solution (4.6) are related to those in solution (5.5) as follows: $c_0 = C_0 (u_0)^{-2}(u_0 - 2\mu)^{\frac{2}{3}}$, $c_1 = C_1$.)

The relation (3.18) reads in this case as follows

$$u^2 \dot{\varphi}^2 = 8\varepsilon \ddot{f}(A - 1) = -\varepsilon c_0 \frac{64\mu(u - 3\mu)}{3(u - 2\mu)^{\frac{5}{3}}} > 0. \tag{5.7}$$

It is satisfied if

$$\varepsilon c_0 < 0 \quad \text{for} \quad u > 3\mu, \tag{5.8}$$

and

$$\varepsilon c_0 > 0 \quad \text{for} \quad 2\mu < u < 3\mu. \tag{5.9}$$

For special case $\varepsilon = 1$ see [19].

This means that for $\varepsilon c_0 < 0$ we have a scalar field function for $u > 3\mu$, i.e. out of the photonic sphere, obeying

$$\frac{d\varphi}{du} = \pm 8 \left(-\varepsilon \frac{c_0 \mu}{3}\right)^{\frac{1}{2}} \frac{(u - 3\mu)^{\frac{1}{2}}}{u(u - 2\mu)^{\frac{5}{6}}}. \tag{5.10}$$

On the contrary, for $\varepsilon c_0 > 0$ we have a scalar field function for $2\mu < u < 3\mu$, i.e. inside the photonic sphere and out of the horizon, obeying

$$\frac{d\varphi}{du} = \pm 8 \left(\frac{\varepsilon c_0 \mu}{3}\right)^{\frac{1}{2}} \frac{(3\mu - u)^{\frac{1}{2}}}{u(u - 2\mu)^{\frac{5}{6}}}. \tag{5.11}$$

We remind that the radius of photonic sphere for Schwarzschild solution in notations under consideration is $3\mu$.

From (3.15), (3.16), (3.17) we obtain

$$E_U = 8u^{-2}(-\mu)(u-2\mu), \tag{5.12}$$
$$F_U = 16u^{-3}\mu(u-3\mu), \tag{5.13}$$
$$G_U = 0, \tag{5.14}$$

and hence using (3.14) and (5.6) we get the following relation for potential function

$$U = U(\varphi(u)) = c_0 \frac{16}{3} u^{-3} \mu (u-3\mu)(u-2\mu)^{-\frac{2}{3}}. \tag{5.15}$$

According to relation (5.8), (5.9) and (5.15) for given $c_0$ we get: $U > 0$ in a domain where $c_0(u-3\mu) > 0$ and $U < 0$ in a domain where $c_0(u-3\mu) < 0$. The equation for scalar field (3.11) (which is equivalent to (2.9)) can be readily verified in this case.

**Solution with "trapped ghost".** Here we consider two solutions: the solution for ordinary scalar field with

$$(A) \quad \varepsilon = 1, \quad u > 3\mu, \tag{5.16}$$

and the solution for phantom scalar field with

$$(B) \quad \varepsilon = -1, \quad 2\mu < u < 3\mu. \tag{5.17}$$

We put without loss of generality

$$c_0 = -1, \quad c_1 = 0. \tag{5.18}$$

We unify these two solutions by using a more general $sEGB$ model with the action $(\kappa^2 = 1)$

$$S = \int d^4z \, |g|^{\frac{1}{2}} \left( \frac{R(g)}{2} - \frac{1}{2} h(\phi) g^{MN} \partial_M \phi \partial_N \phi - \hat{U}(\phi) + \hat{f}(\phi)\mathcal{G} \right). \tag{5.19}$$

Here a new paramerization of our scalar field is used: $\varphi = \varphi(\phi)$ and $h(\phi)$ is a smooth function which satifies the "gluing" condition

$$h(\phi) g^{MN} \partial_M \phi \partial_N \phi = \varepsilon g^{MN} \partial_M \varphi \partial_N \varphi \tag{5.20}$$

for both domains: $(A)$ and $(B)$, and functions $\hat{U}(\phi)$, $\hat{f}(\phi)$ are defined as follows

$$\hat{U}(\phi) = U(\varphi(\phi)), \quad \hat{f}(\phi) = f(\varphi(\phi)). \tag{5.21}$$

For the solutions under consideration the "gluing" relation (5.20) reads

$$h(\phi)\dot{\phi}^2 = \varepsilon \dot{\varphi}^2. \tag{5.22}$$

It may be rewritten by using (5.10) and (5.11) as follows

$$h(\phi(u))\dot{\phi}^2 = \varepsilon 64 \left(\frac{\mu}{3}\right) \frac{|u-3\mu|}{u^2 (u-2\mu)^{\frac{5}{3}}} = 64 \left(\frac{\mu}{3}\right) \frac{(u-3\mu)}{u^2 (u-2\mu)^{\frac{5}{3}}}. \tag{5.23}$$

Here we have used relations $\varepsilon = \text{sign}(u - 3\mu)$ and $\text{sign}(u - 3\mu)|u - 3\mu| = u - 3\mu$, where $\text{sign}(x) = 1$ for $x > 0$ and $\text{sign}(x) = -1$ for $x < 0$.

Now we choose $h(\phi)$ in such way that the solution for $\phi(u)$ has the simplest form:

$$\phi(u) = u. \tag{5.24}$$

Then, we obtain immediately from (5.23) that

$$h(\phi) = 64\left(\frac{\mu}{3}\right) \frac{(\phi - 3\mu)}{\phi^2 (\phi - 2\mu)^{\frac{5}{3}}}. \tag{5.25}$$

The function $h(\phi)$ for $\mu = 1$ is depicted at Figure 1.

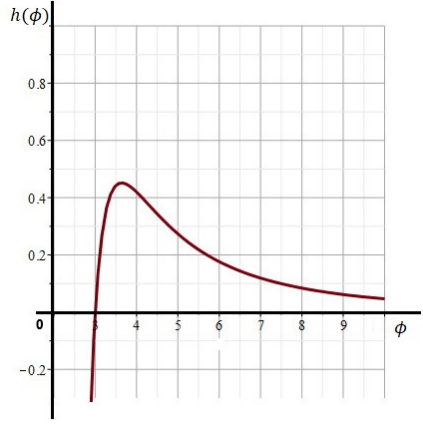

Figure 1: The function $h(\phi)$ from (5.25) for $\mu = 1$.

For the function $h(\phi)$ we have the following asymptotical relations

$$h(\phi) \sim \frac{64\mu}{3} \phi^{-\frac{8}{3}}, \quad \phi \to +\infty, \tag{5.26}$$

$$h(\phi) \sim -\frac{16}{3} (\phi - 2\mu)^{-\frac{5}{3}}, \quad \phi \to 2\mu. \tag{5.27}$$

The function $h(\phi)$ is limited from above. It reaches a maximum at point $\phi_* = (3\sqrt{41} + 39)\mu/16 \approx 3,6381\mu$ with approximate value $h(\phi_*) \approx 0,45182\mu^{-5/3}$. The function $h(\phi)$ tends to $-\infty$ as $\phi \to 2\mu$ and tends to zero as $\phi \to +\infty$.

Analogusly, we find from (5.5) and (5.15) the relations for coupling function and potential

$$\hat{f}(\phi) = -\frac{3}{7} (\phi - 2\mu)^{\frac{1}{3}} \left(\phi^2 + 3\mu\phi + 18\mu^2\right), \tag{5.28}$$

$$\hat{U}(\phi) = -\frac{16}{3} \phi^{-3} \mu (\phi - 3\mu) (\phi - 2\mu)^{-\frac{2}{3}}. \tag{5.29}$$

The coupling function $\hat{f}(\phi)$ from (5.28) for $\mu = 1$ is presented at Figure 2, while the potential $\hat{U}(\phi)$ for $\mu = 1$ is depicted at Figure 3.

The coupling function has the folowing asymptotical behaviours

$$\hat{f}(\phi) \sim -\frac{3}{7}\phi^{\frac{7}{3}}, \quad \phi \to +\infty, \tag{5.30}$$

$$\hat{f}(\phi) \sim -\frac{12}{7}\mu^2 (\phi - 2\mu)^{\frac{1}{3}}, \quad \phi \to 2\mu. \tag{5.31}$$

The function $\hat{f}(\phi)$ is monotonically decreasing from 0 to $-\infty$ ($\phi \in (2\mu, +\infty)$).

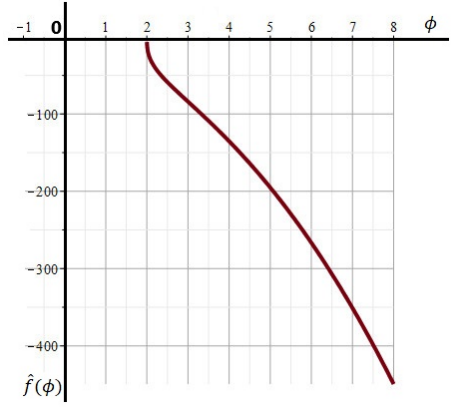

Figure 2: The coupling function $\hat{f}(\phi)$ from (5.28) for $\mu = 1$.

The asymptotical relations for the potential function read

$$\hat{U}(\phi) \sim -\frac{16\mu}{3}\phi^{-\frac{8}{3}}, \quad \phi \to +\infty, \tag{5.32}$$

$$\hat{U}(\phi) \sim \frac{2}{3\mu}(\phi - 2\mu)^{-\frac{2}{3}}, \quad \phi \to 2\mu. \tag{5.33}$$

The function $\hat{U}(\phi)$ is limited from below. It reaches a minimum at point $\phi_* = (3\sqrt{33} + 45)\mu/16 \approx 3,8896\mu$ with approximate value $\hat{U}(\phi_*) \approx 0,052751\mu^{-5/3}$. The function $\hat{U}(\phi)$ tends to $+\infty$ as $\phi \to 2\mu$ and tends to zero as $\phi \to +\infty$.

Here the scalar field is defined in domain

$$\phi > 2\mu. \tag{5.34}$$

For the value $\phi = 3\mu$, corresponding to photonic sphere ($u = 3\mu$), we obtain

$$h(3\mu) = \hat{U}(3\mu) = 0 \tag{5.35}$$

and

$$\hat{f}(3\mu) = -\frac{108}{7}\mu^{\frac{7}{3}}. \tag{5.36}$$

It should be noted that the first (wormhole) solutions with "trapped ghosts" were considered in the paper by Bronnikov and Sushkov [20].

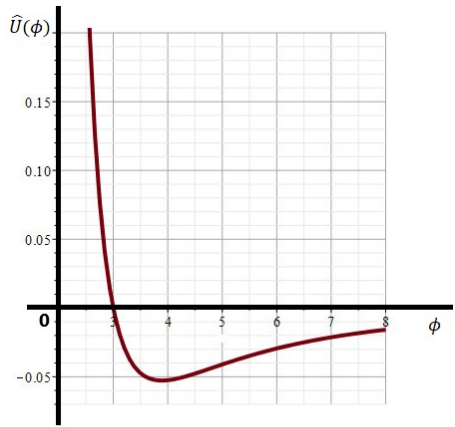

Figure 3: The potential $\hat{U}(\phi)$ from (5.29) for $\mu = 1$.

## 5.2 Ellis wormhole metric

Here we apply the reconstruction procedure to the metric of Ellis wormhole solution (which is a special case of Bronnikov-Ellis wormhole solution [21, 22])

$$ds^2 = -dt^2 + du^2 + \left(u^2 + L^2\right) d\Omega^2. \tag{5.37}$$

Here $L > 0$. This solution appears in the model with minimally coupled phantom scalar field.

The red shift function $A(u)$ and central function $C(u)$ read

$$A(u) = 1, \qquad C(u) = u^2 + L^2. \tag{5.38}$$

The calculation of function $K$ from (3.8) gives us

$$K = \frac{u^2}{u^2 + L^2}. \tag{5.39}$$

For functions $E$, $F$ and $G$ from (3.22), (3.23) and (3.24), respectively, we obtain

$$E = 16\frac{(-L^2)}{u^2 + L^2}, \tag{5.40}$$

$$F = \frac{32uL^2}{\left(u^2 + L^2\right)^2}, \tag{5.41}$$

$$G = 0, \tag{5.42}$$

and hence the master equation (3.21) reads

$$\ddot{f} + \frac{(-2u)}{u^2 + L^2}\dot{f} = 0. \tag{5.43}$$

For $\dot{f} = y$ we have a first order differential equation

$$\frac{dy}{du} = \frac{2u}{u^2 + L^2}y \tag{5.44}$$

with the solution
$$y = \left(u^2 + L^2\right) C_0, \tag{5.45}$$
which implies
$$f = C_0 \left[\frac{u^3}{3} + L^2 u\right] + C_1, \tag{5.46}$$
where $C_0$ and $C_1$ are constants.

The calculation of functions $E_U$, $F_U$ and $G_U$ from (3.15), (3.16) and (3.17) gives us
$$E_U = -\frac{4L^2}{u^2 + L^2}, \tag{5.47}$$
$$F_U = 4\dot{K} = \frac{8L^2 u}{(u^2 + L^2)^2}, \tag{5.48}$$
$$G_U = 0 \tag{5.49}$$

and hence we find from (3.14)
$$U = 0. \tag{5.50}$$

From relation (3.18) we get
$$\varepsilon \dot{\varphi}^2 = -\frac{L^2}{(u^2 + L^2)^2}, \tag{5.51}$$
which urges the scalar field to be phantom one, i.e.
$$\varepsilon = -1, \tag{5.52}$$
and
$$\frac{d\varphi}{du} = \pm \frac{L}{u^2 + L^2}, \tag{5.53}$$
or, equivalently,
$$\varphi - \varphi_0 = \pm \arctan\left(\frac{u}{L}\right), \tag{5.54}$$
where $\varphi_0$ is constant. Reverting this relation we get
$$u = \pm L \tan(\varphi - \varphi_0). \tag{5.55}$$

It follows from (5.46) and (5.55) that
$$f(\varphi) = c_1 + c_0 \left[\tan(\varphi - \varphi_0) + \frac{1}{3} \left(\tan(\varphi - \varphi_0)\right)^3\right], \tag{5.56}$$
where $c_0$, $c_1$ and $\varphi_0$ are constants. (Here $c_0 = \pm C_0 L^3$ and $c_1 = C_1$.)

The scalar field is defined in the domain
$$|\varphi - \varphi_0| < \frac{\pi}{2}. \tag{5.57}$$

We note that
$$f(\varphi_0) = c_1. \tag{5.58}$$

The value $\varphi_0$ appears in the solution for $u = 0$, which corresponds to photonic sphere.

It may be readily verified that the equation for scalar field (3.11) (which is equivalent to (2.9)) is obeyed for this example.

**Remark.** *Without loss of generality one can put $c_0 = 1$, $c_1 = 0$ and $\varphi_0 = 0$. In this case the (reduced) scalar coupling function reads*

$$f(\varphi) = \tan(\varphi) + \frac{1}{3}(\tan(\varphi))^3, \tag{5.59}$$

*where $|\varphi| < \frac{\pi}{2}$. The reduced function $f(\varphi)$ is monotonically increasing from $-\infty$ to $+\infty$ ($\varphi \in (-\pi/2, \pi/2)$). It has the following asymptotical behaviours*

$$f(\varphi) \sim -\frac{1}{3(\varphi \mp \pi/2)^3}, \quad \varphi \to \pm\frac{\pi}{2}, \tag{5.60}$$

$$f(\varphi) \sim \varphi, \quad \varphi \to 0. \tag{5.61}$$

*Graphically the function (5.59) is presented at Figure 4.*

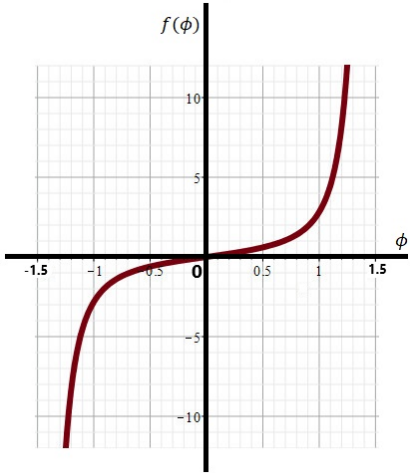

Figure 4: The coupling function $f(\varphi)$ from (5.59).

# 6 Conclusions

Here we have considered the $4d$ gravitational model with scalar field $\varphi$, Einstein and Gauss-Bonnet terms. The action of the model contains potential term $U(\varphi)$, Gauss-Bonnet coupling function $f(\varphi)$ and sign parameter $\varepsilon = \pm 1$: $\varepsilon = 1$ coresponds to ordinary scalar field and $\varepsilon = -1$ - to phantom one.

We have explored a reconstruction procedure for a generic static spherically symmetric metric written in Buchdal parametrisation of radial coordinate $u$: $ds^2 = (A(u))^{-1} du^2 - A(u)dt^2 + C(u)d\Omega^2$, with given $A(u) > 0$ and $C(u) > 0$. For $\varepsilon = 1$ the special case of cenral function $C(u) = u^2$ was considered recently in Refs. [17, 19], while a reconstruction procedure for generic static spherically symmetric metric in another parametrisation of radial coordinate ($r = u$): $ds^2 = (a_1(r))^{-1} dr^2 - a(r)dt^2 + r^2 d\Omega^2$ was studied earlier in Ref. [18] (in presence of fluid matter source).

Along a line as it was done in Ref. [17] we have found relations for $U(\varphi(u))$, $f(\varphi(u))$ and $\frac{d\varphi}{du}$, which give us exact solutions to equations of motion with a given metric. The solutions under consideration are defined up to solutions of master equation, which is a second order linear differential equation for the function $f(\varphi(u))$. We have shown that the solutions to master equation exist for all $A(u) > 0$ and $C(u) > 0$.

We have illustrated the formalism by two examples with: a) Schwarzschild metric and b) Ellis wormhole metric. For $\varepsilon = 1$ the first example a) was studied earlier in Ref. [19]. By gluing two solutions with $\varepsilon = -1$, $2\mu < u < 3\mu$ and $\varepsilon = 1$, $u > 3\mu$ ($2\mu$ is gravitational radius) we have found the solution with a "trapped ghost", which describes an ordinary scalar field outside the photon sphere and phantom scalar field inside the photon sphere.

For the second case b) we have obtained an extension of the Ellis wormhole solution for the sEGB-model with $U(\varphi) = 0$, $\varepsilon = -1$ and coupling function $f(\varphi) = c_1 + c_0(\tan(\varphi) + \frac{1}{3}(\tan(\varphi))^3)$, where $c_1$ and $c_0$ are constants.

It may be worth to devote the forthcoming research to generalization of reconstruction procedure to static higher dimensional spherically symmetric metrics with studying the stability of new solutions obtained in sEGB model by a generalized reconstruction procedure.

## Acknowledgements

One of the authors (VDI) is grateful to K.A. Bronnikov for fruitfull discussions and useful comments. The research was funded by RUDN University, scientific project number FSSF-2023-0003.

# Appendix

## A  The Lagrangian

Here we derive the formula

$$\frac{R(g)}{2} - \frac{1}{2}\varepsilon g^{MN}\partial_M\varphi\partial_N\varphi - U(\varphi) + f(\varphi)\mathcal{G} = L + \frac{dF_*}{du}, \qquad (A.1)$$

for the metric
$$ds^2 = g_{MN}dz^M dz^N = e^{2\gamma(u)}du^2 - e^{2\alpha(u)}dt^2 + e^{2\beta(u)}d\Omega^2, \qquad (A.2)$$

where

$$L = \frac{1}{\kappa^2}\left(e^{\alpha-\gamma+2\beta}\dot{\beta}\left(\dot{\beta}+2\dot{\alpha}\right)+e^{\alpha+\gamma}\right)$$
$$-\frac{1}{2}e^{\alpha-\gamma+2\beta}\varepsilon\dot{\varphi}^2 - e^{\alpha+\gamma+2\beta}U(\varphi) - 8\dot{\alpha}\dot{\varphi}\frac{df}{d\varphi}\left(\dot{\beta}^2 e^{\alpha+2\beta-3\gamma}-e^{\alpha-\gamma}\right), \qquad (A.3)$$

and

$$F_* = -\frac{1}{\kappa^2} \left(\dot{\alpha} + 2\dot{\beta}\right) e^{\alpha-\gamma+2\beta} + f(\varphi) 8\dot{\alpha} \left(\dot{\beta}^2 e^{\alpha+2\beta-3\gamma} - e^{\alpha-\gamma}\right). \tag{A.4}$$

Indeed, the scalar curvature for the metric (A.2) reads

$$R(g) = 2e^{-2\beta} - 2e^{-2\gamma} \left(\dot{\alpha}^2 + 2\dot{\alpha}\dot{\beta} + 3\dot{\beta}^2 - \left(\dot{\alpha} + 2\dot{\beta}\right)\dot{\gamma} + \ddot{\alpha} + 2\ddot{\beta}\right), \tag{A.5}$$

which after multiplication by $|g|^{\frac{1}{2}} = e^{\alpha+\gamma+2\beta}$ gives

$$R(g) |g|^{\frac{1}{2}} = 2e^{\alpha+\gamma} - 2e^{\alpha+2\beta-\gamma} \left(\dot{\alpha}^2 + 2\dot{\alpha}\dot{\beta} + 3\dot{\beta}^2 - \left(\dot{\alpha} + 2\dot{\beta}\right)\dot{\gamma} + \ddot{\alpha} + 2\ddot{\beta}\right). \tag{A.6}$$

Using $|g|^{\frac{1}{2}} = e^{\alpha+\gamma+2\beta}$, we get

$$\frac{R(g)}{2\kappa^2} |g|^{\frac{1}{2}} = \frac{1}{\kappa^2} \left(e^{\alpha-\gamma+2\beta}\dot{\beta}\left(\dot{\beta} + 2\dot{\alpha}\right) + e^{\alpha+\gamma}\right) + \frac{dF_{1*}}{du}, \tag{A.7}$$

where

$$F_{1*} = -\frac{1}{\kappa^2} \left(\dot{\alpha} + 2\dot{\beta}\right) e^{\alpha-\gamma+2\beta}. \tag{A.8}$$

For the Gauss-Bonnet term for the metric (A.2) we obtain

$$\begin{aligned}\mathcal{G} = &-8e^{-2\beta-2\gamma}\dot{\alpha}^2 + 8e^{-4\gamma}\dot{\alpha}^2\dot{\beta}^2 + 16e^{-4\gamma}\dot{\alpha}\dot{\beta}^3 + 8e^{-2\beta-2\gamma}\dot{\alpha}\dot{\gamma} \\ &-24e^{-4\gamma}\dot{\alpha}\dot{\beta}^2\dot{\gamma} - 8e^{-2\gamma-2\beta}\ddot{\alpha} + 8e^{-4\gamma}\dot{\beta}^2\ddot{\alpha} + 16e^{-4\gamma}\dot{\alpha}\dot{\beta}\ddot{\beta},\end{aligned} \tag{A.9}$$

which after multiplication by $|g|^{\frac{1}{2}} = e^{\alpha+\gamma+2\beta}$ gives the total derivative

$$\begin{aligned}\mathcal{G} |g|^{\frac{1}{2}} = &-8e^{\alpha-\gamma}\dot{\alpha}^2 + 8e^{\alpha+2\beta-3\gamma}\dot{\alpha}^2\dot{\beta}^2 + 16e^{\alpha+2\beta-3\gamma}\dot{\alpha}\dot{\beta}^3 + 8e^{\alpha-\gamma}\dot{\alpha}\dot{\gamma} \\ &-24e^{\alpha+2\beta-3\gamma}\dot{\alpha}\dot{\beta}^2\dot{\gamma} - 8e^{\alpha-\gamma}\ddot{\alpha} + 8e^{\alpha+2\beta-3\gamma}\dot{\beta}^2\ddot{\alpha} + 16e^{\alpha+2\beta-3\gamma}\dot{\alpha}\dot{\beta}\ddot{\beta} \\ = &\frac{d}{du}\left(8\dot{\alpha}\left(\dot{\beta}^2 e^{\alpha+2\beta-3\gamma} - e^{\alpha-\gamma}\right)\right),\end{aligned} \tag{A.10}$$

and hence

$$f(\varphi) |g|^{\frac{1}{2}} \mathcal{G} = -8\dot{\alpha}\dot{\varphi}\frac{df}{d\varphi} \left(\dot{\beta}^2 e^{\alpha+2\beta-3\gamma} - e^{\alpha-\gamma}\right) + \frac{dF_{2*}}{du}, \tag{A.11}$$

where

$$F_{2*} = 8f(\varphi)\dot{\alpha}\left(\dot{\beta}^2 e^{\alpha+2\beta-3\gamma} - e^{\alpha-\gamma}\right). \tag{A.12}$$

Using $g^{MN}\partial_M\varphi\partial_N\varphi = g^{uu}\dot{\varphi}^2 = e^{-2\gamma}\dot{\varphi}^2$ and formulas (A.7), (A.11) we get relations (A.1), (A.3), (A.4). Here $F_* = F_{1*} + F_{2*}$.